\documentclass[aps,prb,twocolumn]{revtex4}
\usepackage{graphicx}
\usepackage{dcolumn}
\usepackage{amsmath}
\usepackage{array}
\usepackage{epsfig}

\begin{document}

\title{Doping dependence of the lattice dynamics in Ba(Fe$_{1-x}$Co$_x$)$_2$As$_2$ studied by Raman spectroscopy}
\author{L. Chauvi\`ere}
\author{Y. Gallais}
\author{M. Cazayous}
\author{A. Sacuto}
\author{M.A. M\'easson}
\affiliation{Laboratoire Mat\'eriaux et Ph\'enom\`enes Quantiques, UMR 7162 CNRS, Universit\'e Paris Diderot , B$\hat{a}$t. Condorcet 75205 Paris Cedex 13, France}
\author{D. Colson}
\author{A. Forget}
\affiliation{Service de Physique de l'Etat Condens\'e, DSM/DRECAM/SPEC, CEA Saclay, 91191 Gif-sur-Yvette, France}

\begin{abstract}
We report Raman scattering measurements on iron-pnictide superconductor Ba(Fe$_{1-x}$Co$_x$)$_2$As$_2$ single crystals with varying cobalt $x$ content. Upon cooling through the tetragonal-to-orthorhombic transition, we observe a large splitting of the E$_g$ in-plane phonons modes involving Fe and As displacements. The splitting of the in-plane phonons at the transition is strongly reduced upon doping and disappears for $x=0.06$ qualitatively following the trend displayed by the Fe magnetic moment. The origin of the splitting is discussed in terms of magnetic frustration inherent to iron-pnictide systems and we argue that such enhanced splitting may be linked to strong spin-phonon coupling. 
 
\end{abstract}

\maketitle

\section{INTRODUCTION}
The discovery of superconductivity in the compound LaO$_{1-x}$F$_x$FeAs with a maximal $T_c$ of 26~K \cite{Kamihara} has led to a great excitement in the scientific community. Since then, an entirely new family of high-$T_c$ superconductors based on iron-arsenide superconducting layers has been discovered. Two main groups of iron-pnictides can be distinguished : the ROMP (R = Rare-earth, O = Oxygen, M = transition Metal, P = Pnictogen) or 1111 family, and the AM$_2$P$_2$ (A = earth-Alcaline) or 122 family. All these compounds share the same building blocks, FeAs planes where Fe and As are in tetrahedrical coordination. There are strong evidences that the FeAs planes control the transport, magnetic and superconducting properties of these systems \cite{Schmalian}. The undoped compounds display magnetic order of Spin Density Wave (SDW) type at low temperature. They are not Mott insulators like undoped cuprates but rather bad metals with high resistivities. Upon doping with electrons or holes, or in some compounds by applying hydrostatic pressure, the magnetic order and its associated structural transition are suppressed and a superconducting region emerges. Upon doping, $T_c$ can be raised until 55 K in SmO$_{0.8}$F$_{0.1}$FeAs \cite{Ren} even 56 K in Gd$_{0.8}$Th$_{0.2}$OFeAs \cite{Wang}.

\par
While it is clear that the suppression of the magnetic/structural transition is a requisite to the emergence of the superconducting order, there is no clear picture yet of the interplay between lattice and magnetic degrees of freedom. In particular, the relationship between the structural and the magnetic transitions has been discussed extensively reflecting contrasting approaches to the physics of the pnictides. In the strongly correlated or  Mott-Hubbard approaches, it has been suggested that the magnetic and structural transitions are intimately connected due to competing super-exchange interactions in the Fe-plane which lead to magnetic frustration\cite{Si,Yildirim}. This frustration is believed to explain the rather low Fe magnetic moment. In this local moment scenario, the orthorhombic distortion removes the magnetic frustration of the underlying tetragonal lattice. Another class of approaches emphasizes the itinerant character of electrons in these systems and rather link the transition to Fermi surface nesting \cite{Schmalian}. The itinerant picture is driven by the relative success of Density Functional Theory (DFT) calculations to predict the correct magnetic order, in stark contrast to the case of cuprates. It is also supported by the relative agreement between the calculated band-structure and ARPES experiments\cite{Ding}. The possibility of strong coupling between magnetic and structural degrees of freedom is especially relevant since it could give clues to the underlying mechanism of high-temperature superconductivity in the iron-pnictides.

\par
Here, we report a Raman scattering study of the lattice dynamics as a function of electron doping in Ba(Fe$_{1-x}$Co$_x$)$_2$As$_2$. The impact of the structural and magnetic transition on zone-center phonons is investigated via temperature dependent measurement of the Raman phonon modes across the transition. Several anomalies are detected in the Fe or As related modes. The most salient one is a very large splitting of the doubly degenerate in-plane Fe-As mode which occurs at the tetrahedral to orthorhombic transition. The amplitude of the splitting, about 9~cm$^{-1}$ for the undoped compound, is too large to be explained solely by the weak orthorhombic distortion and could be linked to strong spin-phonon coupling.

\section{EXPERIMENTAL DETAILS}
We have focused our study on the double layered compound, of the 122 family Ba(Fe$_{1-x}$Co$_x$)$_2$As$_2$ which is electron-doped with Cobalt inserted directly in the Fe planes. Previous experiments have determined its phase diagram using different experimental techniques such as resistivity, heat capacity, magnetic susceptibility \cite{Chu,Colson}, neutron scattering and X-ray diffraction measurements \cite{Huang,Lester,Pratt}. This compound exibits superconductivity between $x = 0.025$ and $x = 0.18$ doping with a maximal $T_c = 24~K$ for $x = 0.07$ doping. At low doping, it undergoes a magnetic transition from a non-magnetic state to an antiferromagnetic long-range order associated with a Spin-Density-Wave (SDW). In the magnetically ordered phase, the Fe spins are aligned antiferromagnetically along $a$-axis and ferromagnetically along $b$-axis ($a>b$) in a stripe-like pattern. Very close to the magnetic transition, a structural one occurs from a tetragonal phase at high temperature to an orthorhombic one at low temperature. The two transitions are simultaneous for undoped Ba-122 while they split upon Co doping, the magnetic transition occurring at slightly lower temperatures \cite{Chu,Lester}. The small orthorhombic distortion of the crystal structure with cooling changes the lattice parameters of the Fe-As planes and induces different bond lengths along $a$ and $b$-axis \cite{Huang}. 

\begin{figure}
	\centering
	\epsfig{figure=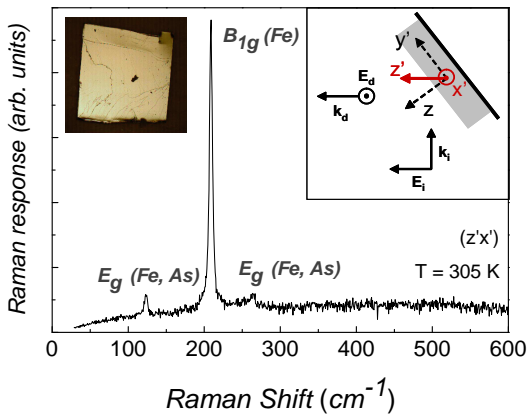, width=0.8\linewidth, clip=}
	\caption{Raman spectrum of BaFe$_2$As$_2$ at 305 K in the \textit{(z'x')} configuration at $\lambda = 514.52 nm$. Inset :  optical image of the undoped single crystal BaFe$_2$As$_2$ - geometry of the \textit{(z'x')} configuration.}
	\label{fig1}
\end{figure}

\par
Single crystals of Ba(Fe$_{1-x}$Co$_x$)$_2$As$_2$ were grown from a self-flux method by high-temperature solid-state reactions as described elsewhere \cite{Colson}. Typical crystal sizes are $2\times2\times0.1~mm^3$. Samples were cleaved to perform experiments on high-quality surfaces as shown in the inset of Fig. \ref{fig1}. Results on six different single crystals with Co doping $x=0$, $x=0.02$, $x=0.03$, $x=0.04$, $x=0.045$ and $x=0.06$ are reported in this study. Their superconducting transition temperatures were measured by SQUID magnetometry and by in-plane transport \cite{Colson}. Freshly cleaved single crystals were held in a vacuum of $10^{-6}~mbar$  and cooled by a closed-cycle refrigerator. All the spectra reported here were performed using the $\lambda = 514.52~nm$ line of an Argon-Krypton laser. Typical laser power densities focused on the sample range from 20 to 200~W/cm$^2$ (the lower power density was used for the lowest temperature spectra). We estimated the laser heating of the sample by comparing Stokes and anti-Stokes Raman spectra and also via the evolution of phonon frequencies with incident laser power at constant cold finger temperature. All the temperatures reported here take into account the estimated laser heating. The scattered light was analyzed by a triple grating spectrometer (JY - T64000) equipped with a liquid nitrogen cooled CCD detector. Typical Raman phonon intensities were very low and long acquisition times were needed (typically 30~minutes). Incident light was polarized along \textit{z'} direction (i.e. with a finite projection along the $c$-axis or [001] crystallographic direction) or along the \textit{x'} direction (i.e. along the [110] crystallographic direction) with quasi-Brewster incidence as shown in the inset of Fig. \ref{fig1} . Scattered light polarization was collected along $x'$ or $z'$ directions. Thus four different configurations, \textit{(z'x')} - \textit{(z'z')} - \textit{(x'z')} - \textit{(x'x')}, were obtained by combining the different incident and scattered photon polarizations. We note that photons polarized along the \textit{z'} direction can probe $ab$-plane polarized phonons because of the finite projection of their polarization along the $c$-axis. 

\begin{table}
	\centering
	\begin{tabular}{|c|c|c|c|c|c|}
		\hline
		phonon frequency & \textit{(z'x')} & \textit{(z'z')} & \textit{(x'z')} & \textit{(x'x')} & phonon symmetry \\
		\hline
		$124~cm^{-1}$ & O & X & X & X & $E_{g}(Fe, As)$ \\
		\hline
		$209~cm^{-1}$ & O & X & O & X & $B_{1g}(Fe)$ \\
		\hline
		$264~cm^{-1}$ & O & X & X & X & $E_{g}(Fe, As)$ \\
		\hline
	\end{tabular}
\caption{Raman peak frequencies, polarisation rules and phonon symmetry.}
\label{polar}
\end{table}

\section{RESULTS}
Figure \ref{fig1} shows a typical Raman spectrum of BaFe$_2$As$_2$ in the \textit{z'x'} configuration at ambient temperature. The phonon modes display typical Lorentzian lineshapes with no detectable asymmetry. Table \ref{polar} summarizes the polarisation rules of the involved phonons. The undoped single crystal BaFe$_2$As$_2$ has a tetragonal symmetry ($I4/mmm$) at room temperature. From symmetry considerations, one expects four Raman active phonons A$_{1g}(As)$, B$_{1g}(Fe)$, E$_{g}(Fe, As)$, E$_{g}(Fe, As)$. Considering the polarisation rules given in Table \ref{polar} and following Litvinchuk et al.  \cite{Litvinchuk}, we indexed the phonons peaks as 209~$cm^{-1}$ for B$_{1g}$ mode, which is a pure mode involving displacement of Fe-atoms along the $c$-axis, 124~$cm^{-1}$ for the low-frequency E$_{g}$ mode and 264~$cm^{-1}$ for the high-frequency E$_{g}$ mode, which are strongly mixed modes, involving displacements of both Fe and As atoms in the $ab$-planes. We did not observe the Arsenic mode A$_{1g}(As)$ at room temperature. 

\begin{figure}
	\centering
	\epsfig{figure=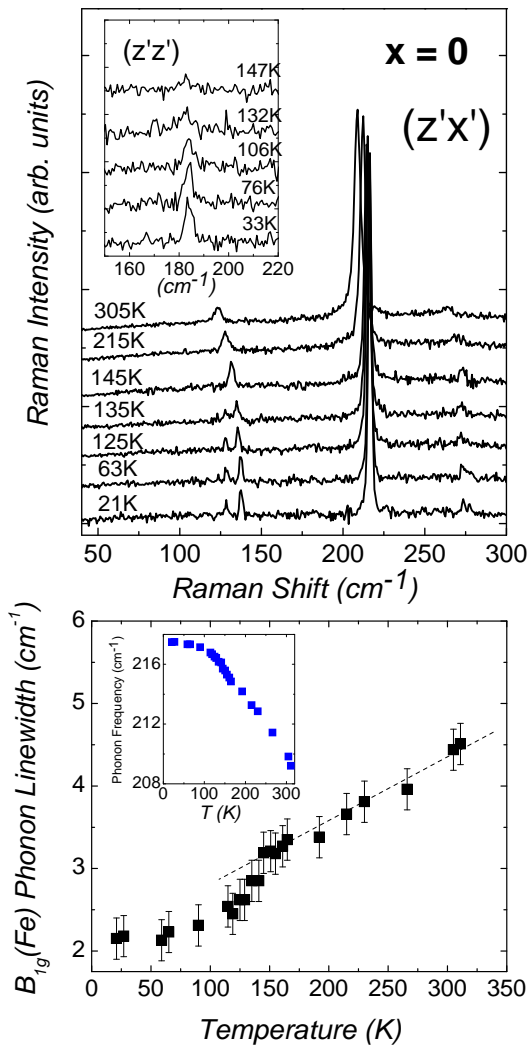, width=0.8\linewidth, clip=}
	\caption{Top panel : evolution with temperature of the Raman spectrum of $BaFe_2As_2$ in the \textit{(z'x')} configuration. Inset : zoom on the $A_{1g}(As)$ mode. Bottom panel : evolution of the B$_{1g}$ Fe phonon linewidth with temperature. Inset : evolution of the B$_{1g}$ Fe phonon frequency with temperature.}
	\label{fig2}
\end{figure}

\par
The evolution of the Raman spectrum of undoped BaFe$_2$As$_2$ with temperature in \textit{(z'x')} configuration is shown in Fig. \ref{fig2}. The spectra were vertically shifted for clarity. While the B$_{1g}(Fe)$ mode simply hardens with cooling, a clear splitting of the low-frequency E$_{g}$ phonon is observed between 135~K and 145~K $\pm$ 5 K. The temperature range is consistent with the magneto-structural transition temperature previously reported in Ba-122 \cite{Huang}. The high-frequency E$_{g}$ mode is too weak to assert its splitting but the lowest temperature spectrum is indicative of a similar albeit smaller splitting. As shown in the inset, the A$_{1g}(As)$ mode, undetectable at room temperature, gains considerably in intensity in the same temperature range (135~K - 145~K), i.e. across the phase transition. The considerable impact of the phase transition on the As phonon intensity was also observed on Ca-122 by Choi et al.\cite{Choi}. We note that such anomaly in the As phonon intensity is not found in 1111 systems \cite{Hadjiev, Gallais}.

\par
Upon closer inspection, the B$_{1g}(Fe)$ mode also displays an anomaly in its linewidth at the transition as shown in the lower panel of Fig. \ref{fig2}. Contrary to Ca-122, the phonon frequency does not exhibit any anomaly but its linewidth shows a sizable decrease across the transition. In an electron-phonon coupling model, this effect can be ascribed to the opening of a pseudogap in the electronic spectrum associated with SDW formation \cite{Choi}. Such a pseudogap is indeed observed below 500~cm$^{-1}$ in the far-infrared conductivity spectrum of Ba-122 by Pfuner et al. \cite{Pfuner}. In the following, we will argue however that spin-phonon coupling might play also a role in this anomaly and thus provide an alternative scenario. 

\par
The most striking feature of the data presented in Fig. \ref{fig2} is the very large splitting of the low energy E${_g}$ phonon across the phase transition. At first sight, the splitting of the doubly degenerate in-plane E${_g}$ phonons is consistent with the orthorhombic distortion that occurs at the structural transition. At the transition, the lattice parameters in the $ab$-plane change ($a>b$) and the degeneracy of the atomic displacements involved in the E${_g}$ phonon is lifted. However, the amplitude of the splitting, around 7~$\%$ of the mode energy, is anomalously large. In fact, such a splitting has been already discussed and thought to be to small to be resolved \cite{Litvinchuk,Yildirim}. Indeed a simple approach linking  the phonon frequency $\omega$ with the bond length $l$ ($\omega^2 \sim \frac{1}{l^3}$) \cite{Martin} would yield a splitting of less than 1~cm$^{-1}$ based on crystallographic data \cite{Huang} and therefore cannot explain a phonon splitting of around 9~$cm^{-1}$.   

\begin{figure}
	\centering
	\epsfig{figure=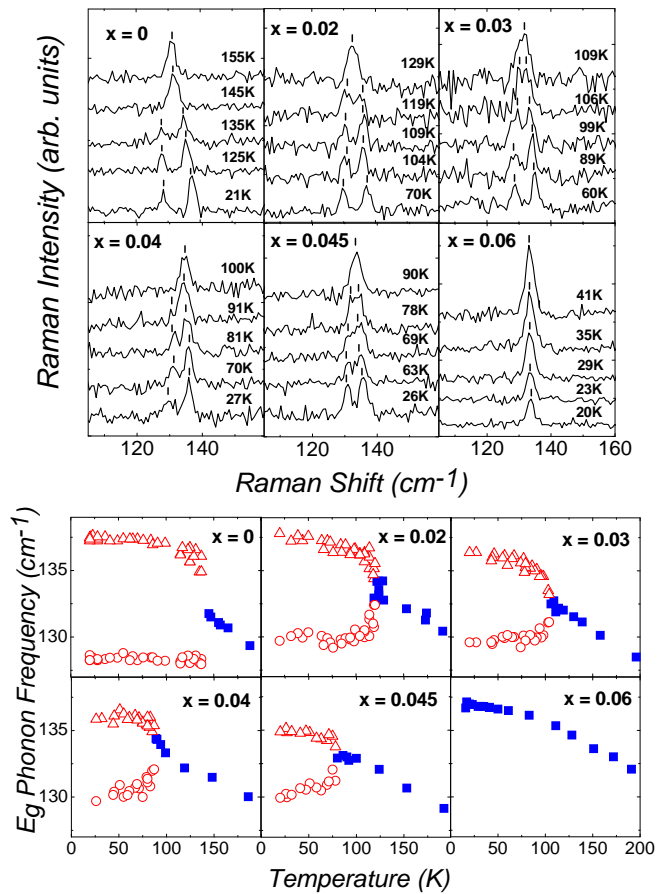,width=0.99\linewidth,clip=}
	\caption{Top panel : evolution with temperature of the Raman spectrum of Ba(Fe$_{1-x}$Co$_x$)$_2$As$_2$ in the \textit{z'x'} configuration for $x = 0$, $x = 0.02$, $x = 0.03$, $x = 0.04$, $x = 0.045$, $x = 0.06$. Bottom : evolution of the E$_{g}(Fe, As)$ phonon frequency with temperature for the same dopings (on bottom).}
	\label{fig3}
\end{figure}

\par
\begin{figure}
	\centering
	\epsfig{figure=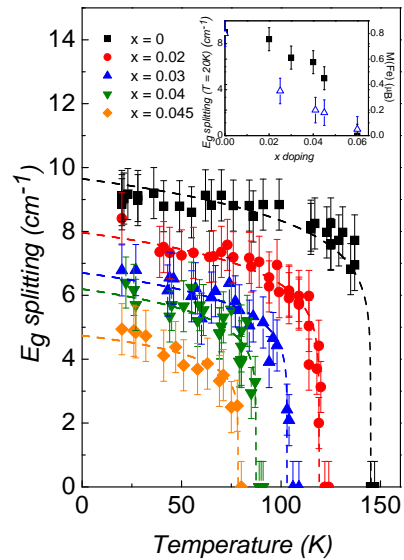, width=0.6\linewidth, clip=}
	\caption{Evolution of the amplitude of the E$_g$ phonon splitting with temperature for different doping $x = 0$, $x = 0.02$, $x = 0.03$, $x = 0.04$, $x = 0.045$. The dashed lines are fits using the order parameter power law described in the text using $\beta=0,12$. Inset : evolution of the amplitude of the splitting at low temperature (20~K) and of the magnetic moment of Fe with $x$ doping\cite{Huang,Lester,Laplace}.}
	\label{fig4}
\end{figure}

The $Co$ doping dependence of the low energy E$_g$ phonon splitting is shown in Fig. \ref{fig3}, where for each doping we report the evolution of the $E_{g}$ phonon frequency with temperature. As the $Co$ doping increases, the splitting occurs at lower temperature following the trend observed for the structural transition. For $x=0.06$, no splitting is observed for temperatures as low as 20~K in agreement with neutron and specific heat data which did not detect any structural phase transition for similar doping \cite{Lester,Chu}. We note however that recent NMR and resistivity measurements performed on similar crystals with $x=0.06$ were able to detect a SDW transition slightly above 30~K\cite{Colson, Laplace}. The extracted Fe magnetic moment from NMR data was estimated to be very weak however : 0.05~$\mu_B$ instead of 0.9~$\mu_B$ for the undoped crystals \cite{Lester}. Our data are summarized in Fig. \ref{fig4} where the amplitude of the splitting is reported as a function of both doping and temperature. Electron doping via Co doping reduces both the transition temperature and the amplitude of the splitting measured at 20~K. It decreases from 9~cm$^{-1}$ at $x=0$ to 5~cm$^{-1}$ at $x=0.045$. 

\par
\begin{figure}
	\centering
	\epsfig{figure=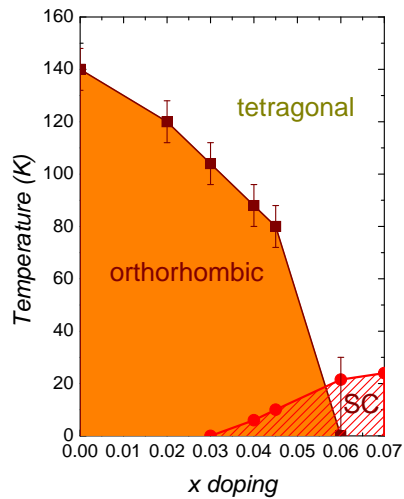, width=0.6\linewidth, clip=}
	\caption{Phase diagram of $Ba(Fe_{1-x}Co_x)_2As_2$ : temperatures of the structural (T$_s$) and the superconducting (T$_c$) transitions versus $x$ doping (Co content).}
	\label{fig5}
\end{figure}

The amplitude of the phonon splitting, being directly linked to the order parameter of the tetragonal-orthorhombic structural transition, gives us important information on the nature of the transition. Following recent neutron scattering measurements \cite{Wilson} on undoped Ba-122 showing a continuous transition for both the magnetic and the structural order parameters, we have performed an order parameter analysis, $\Delta\omega(T)=\Delta\omega(T=0~K)\times(1-\frac{T}{T_s})^\beta$ where $\Delta\omega$ is the phonon splitting, T$_s$ the transition temperature and $\beta$ the critical exponent. Within our experimental accuracy, $\beta$ values ranging 0.1 and 0.2 were found to reproduce satisfactorily the temperature dependence of all doping levels, suggesting a second order-like transition at all dopings in agreement with neutron data of Wilson et al. \cite{Wilson}.  
The temperature of the structural transition $T_s$ was extracted from the order parameter analysis and is summarized in Fig.  \ref{fig5}. The evolution of $T_s$ with Co doping is in overall agreement with neutron scattering data of Lester et al. \cite{Lester}. We have added on the same figure the superconducting transition temperatures $T_c$ of the samples measured with a SQUID magnetometer. We note that the structural transition is still present in the range of Co doping where superconductivity emerges in agreement with neutron, transport and NMR data \cite{Lester,Colson,Laplace}.

\begin{figure}
	\centering
	\epsfig{figure=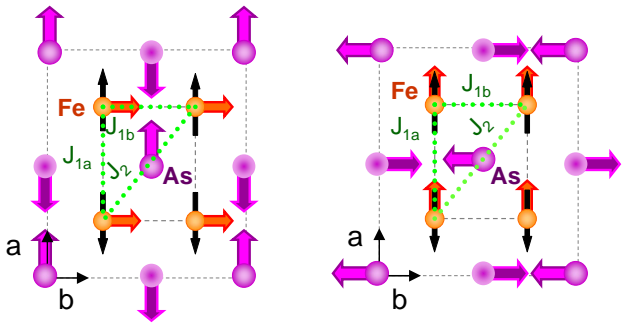, width=0.95\linewidth, clip=}
	\caption{In plane Fe-As phonon displacement patterns in the orthorhombic phase with the positions of the Fe and As atoms projected in the $ab$ plane \cite{Litvinchuk,Zbiri}. The modes arise from the doubly degenerate E$_{g}$ mode in the tetragonal structure ($I4/mmm$) which splits into non-degenerate B$_{2g}$ and B$_{3g}$ modes in the orthorhombic structure ($Fmmm$). For As atoms, full and empty symbols are meant to differentiate between the As atoms located above and below the Fe plane. The Fe-spins are drawn in black and display the stripe-like magnetic pattern observed by neutron scattering in the orthorhombic phase \cite{Huang}. The short range exchange integrals involving next-neighbors (J$_{1a}$ and J$_{1b}$) and nearest next neighbors (J$_2$) Fe spins are also depicted. Note that the superexchange path of J$_{1a}$ and J$_{1b}$ involves the As atoms.}
	\label{fig6}
\end{figure}

\section{DISCUSSION}
As stated earlier, while the E$_g$ phonon splitting is directly linked to the structural transition, the measured orthorhombic distortion for $x=0$ cannot account solely for the large splitting amplitude observed here. This very large phonon splitting leads us to suspect that magnetic ordering has a strong impact on lattice dynamics and may enhance the splitting significantly. As previously emphasized, it is widely believed that the magnetic and structural degrees of freedom are strongly connected in iron-pnictides. For example, ab-initio calculations show that the lattice dynamics depend strongly on both the Fe-spin state and the exact magnetic ordering. In particular, better agreement is found between the phonon density of state observed by neutron and X-ray scattering and ab-initio calculations when the magnetic ordering is taken into account \cite{Yildirim2}. This view is reinforced by the observation that the doping dependence of the low temperature amplitude of the phonon splitting follows a similar trend as the one reported for the Fe magnetic moment as shown in the inset of Fig. \ref{fig4}. Neutron scattering measurements by Wilson et al. \cite{Wilson} also show a direct correlation between the temperature dependence of the magnetic and the structural order parameters in undoped BaFe$_2$As$_2$.   

\par
A simple view of the magnetic order based on localized Fe spins provides an intuitive picture of the interplay between magnetic and structural degrees of freedom in iron-pnictides. The dominant exchange path between Fe spins are along the  $a$, $b$ axis and along the diagonals via As atoms \cite{Si}. In the tetragonal phase, all the exchange paths are dominantly antiferromagnetic yielding magnetic frustration. The frustration is lifted in the orthorhombic phase where the exchange integrals along the $a$ and $b$ axis become unequivalent with a ferro and antiferro spin alignment along the $b$ and $a$ axis respectively (see Fig. \ref{fig6}).  Assuming a strong spin-phonon coupling, the large difference in exchange integrals along $a$ and $b$ directions  \cite{Yildirim2}, J$_{1a}$ and J$_{1b}$ (see figure \ref{fig6}),  axis could in principle explain the large splitting of the in-plane E$_g$ phonon \cite{Lockwood}. A similar scenario has been proposed in frustrated Heisenberg antiferromagnets on a pyrochlore lattice where large phonon splitting have been observed by infrared spectroscopy across the magneto-structural transition \cite{Sushkov}. In frustrated magnetic systems, the strong spin-phonon coupling is due to a modulation of the exchange integrals by the lattice mode displacements involving magnetic ions. In these systems, the magnetic degrees of freedom usually drive the structural distortion, a scenario akin to the spin-Peierls transition \cite{Yamashita, Sondhi}. In our case the non-degenerate phonon modes, as shown in Fig. \ref{fig6}, are mixed Fe-As modes and their displacement patterns modulate the Fe-As-Fe bond angles and thus all the short range exchange integrals between Fe atoms which occur mainly via the As orbitals. In particular, it is clear that the displacement pattern of each mode distorts differently the Fe-As-Fe angle involved in the J$_{1a}$ and J$_{1b}$ exchange integrals respectively. Such spin-phonon interaction is expected to provide an additional contribution to the splitting of the E$_{g}$ phonon mode and may also explain the dramatic enhancement of the $As$ phonon intensity below T$_s$ \cite{Choi,Lockwood, Lockwood2}. 

\par
While qualitatively consistent with our data, the validity of such a localized spin scenario in itinerant systems such as iron-pnictides is however debatable \cite{Mazin}. In addition in the spin-Peierls-like picture presented here, the phonon splitting is expected to scale with the square of the magnetic moment \cite{Lockwood} in disagreement with both our data and neutron scattering measurements \cite{Wilson}. An itinerant approach linking the phonon displacement patterns to nesting properties of the Fermi surface might be more suitable to explain the large phonon splitting. Recent DFT calculations show a strong impact of Raman active phonons on the electronic density of state near the Fermi level and in particular for the E$_g$ mode discussed here \cite{Zbiri}. Clearly more work is needed to understand the coupling between spins and phonons degrees of freedom in itinerant systems such as the iron-pnictides.

\section{CONCLUSION}
In conclusion, we have reported a doping dependent Raman scattering study of the lattice dynamics in Ba(Fe$_{1-x}$Co$_x$)$_2$As$_2$. A large splitting of the in-plane Fe-As phonon across the tetragonal-orthorhombic structural transition has been observed. The phonon splitting is found to weaken upon Co doping and disappears for $x=0.06$. The temperature dependence of the splitting below transition temperature is consistent with a continuous or second order transition. The amplitude of the phonon splitting cannot be accounted by the orthorhombic distortion alone and might be a fingerprint of strong spin-phonon coupling in iron-pnicides.

\end{document}